\documentclass{article}
\usepackage{spconf,amsmath,graphicx}
\usepackage{multirow}
\usepackage{subcaption}
\usepackage{caption}
\usepackage{xcolor}
\usepackage{soul}
\usepackage{amsfonts} 

\usepackage{algorithm}
\usepackage{algpseudocode}

\usepackage{hyperref}       
\usepackage{booktabs}       

\title{A Lightweight dynamic filter for keyword spotting}
\name{Donghyeon Kim$^1$, Kyungdeuk ko$^1$, Jeonggi Kwak$^1$, David K. Han$^2$, Hanseok Ko$^1$\thanks{This work was supported by Korea Environment Industry \& Technology Institute(KEITI) through Exotic Invasive Species Management Program, funded by Korea Ministry of Environment(MOE) (2021002280004)}\thanks{David Han's work on this paper is partly sponsored by the Office of Naval Research (.Grant Number: N00014-21-1-2790.).}}
\address{
  $^1$Korea University, South Korea\\
  $^2$Drexel university, USA}

\begin{document}
\ninept
\maketitle
\begin{abstract}
Keyword Spotting (KWS) from speech signals is widely applied to perform fully hands-free speech recognition. The KWS network is designed as a small-footprint model so it can continuously be active. Recent efforts have explored dynamic filter-based models in deep learning frameworks to enhance the system's robustness or accuracy. However, as a dynamic filter framework requires high computational costs, the implementation is limited to the computational condition of the device. In this paper, we propose a lightweight dynamic filter to improve the performance of KWS. Our proposed model divides the dynamic filter into two branches to reduce computational complexity: pixel level and instance level. The proposed lightweight dynamic filter is applied to the front end of KWS to enhance the separability of the input data. The experimental results show that our model is robustly working on unseen noise and small training data environments by using a small computational resource.
\end{abstract}

\section{Introduction}
Recently, deep learning based speech applications have been applied in real world applications \cite{8933025,9466122,kim2021specmix,kim2021multi}.
For smart devices on standby for human user's commands, Keyword Spotting (KWS) is an essential capability. In addition, KWS system can be leveraged for Over-The-Top (OTT) media service or smart TV to enhance accessibility between user and computer. As these systems listen continuously to the audio stream in the environment, their KWS process should cost minimally to conserve battery power. For this reason, model parameters and subsequent computational cost are important elements in evaluating the KWS system.
This 
Conventional KWS has been based on Hidden Markov Models (HMMs) by using a large vocabulary speech recognizer and a background model \cite{benayed2003confidence,ketabdar2006posterior}. 
With the advent of deep learning framework, Multi-Layer Perceptron (MLP) and Convolutional Neural Network (CNN) have been shown to outperform the conventional methods \cite{chen2014small,sainath2015convolutional,arik2017convolutional}.
As CNN requires high computations in general to extract spectral and temporal features, some proposed models focused to reduce computational cost. Depthwise Separable Convolution (DSConv) extracts channel-wise features separately and later combines channel information by using 1D convolution \cite{zhang2017hello}. As an alternative to the low computational model, temporal convolution-based networks \cite{choi2019temporal,li2020small} have been proposed to extract spectral features by performing convolution on the temporal axis.
Neural Architecture Search (NAS) methods \cite{mo2020neural,zhang2021autokws} have been also proposed to find the best network structure with a reduced computational requirement for the KWS model. 
Although these networks would reduce memory footprints and computational operations, their performances have been shown limited, particularly with unknown speakers or unseen noise. 

An alternative to improve performance is integrating a filter on the front end of the process \cite{jia2016dynamic,kim2020dual}. In the conventional dynamic filter process, dynamic weights are generated by each input patch. In this case, computations would be required for all the input pixels. Such an extra process of filtering, however, adds to the computational cost, which defeats the purpose of reducing computations. 
To balance out the requirement between high performance with minimal computation, we propose a Lightweight Dynamic (LDy) convolution for the front-end filtering. In our proposed network, two branches of dynamic filters are implemented at the front end of the network: pixel-level and instance-level.  A pixel-level dynamic filter performs pixel attention for Time-Frequency (T-F) features while an instance-level dynamic filter produces a global representative weight vector from the temporal pooling of the acoustic features. These two ways of dynamic weights are merged to conduct CNN process. We also leverage dynamic weights to conduct Instance normalization to the output of dynamic convolution. This is because conventional normalization methods \cite{ba2016layer,ulyanov2016instance,chang2021subspectral}, which use static weights, have limitations in addressing the noise robustness problem. Our proposed model is applied to the convolution process in a computationally efficient manner to deliver dynamic filtering robust to noisy environments.
The KWS experiments are carried out on Speech command data \cite{warden2018speech} v1 and v2. Additionally, we utilize three different noise data \cite{mesaros2019acoustic,salamon2014dataset,Wichern2019WHAM}, to set up the unseen noise environments. The experimental results show that the proposed lightweight dynamic filter improves KWS performance with robustness over recently developed methods. Especially our lightweight dynamic convolution only utilizes 220K Flops., and 2K parameters for its implementation.

\section{Related Works}

Dynamic Filter Network (DFN) is an adaptive deep neural network architecture \cite{jia2016dynamic} that generates filter parameters by neural networks. DFN consists of a filter generator and dynamic filter layers. The filter generator produces weights of the neural network, and the process of filtering occurs in the dynamic filter layer with the generated weights. This method conditionally adjusts weight vectors for given patch-level features. This learning method is simpler and works more efficiently over a self-attention network \cite{wu2019pay}.  
Kim \textit{et al.} \cite{kim2020dual} proposed a convolution-based dynamic filter network to enhance salient features from the unseen noisy audio stream, and their experimental results showed that their approach outperformed conventional feature enhancement methods.
Fujita \textit{et al.} \cite{fujita2020attention} also utilized a dynamic filter-based method for a lightweight ASR model. They confirmed that applying dynamic convolution to the decoder part in the encoder-decoder model improves accuracy and reduces computational load over transformer\cite{vaswani2017attention} based models. 
Although dynamic filter-based approaches have shown progressive performance improvements recently over other methods, they typically require high computational cost in filter implementation since the weights are produced by each patch basis. To alleviate this issue, we borrow the idea of Decoupled Dynamic Filter (DDF) \cite{zhou2021decoupled} from computer vision. Instead of a single filter to take up channel-spatial features directly, DDF divides the filtering into two parts: channel and spatial. The channel-wise filter applies computationally efficient 1-D convolution while the spatial-wise side uses a simple global average pooling. These two filters are trained at separate branches of the network and they are then combined into forming a dynamic filter for input feature maps. By using simple computations, this method yields a significant reduction in memory and floating-point operations (FLOPs). In our adaptation of DDF, we split the dynamic filter process into Pixel Dynamic Filter (PDF) and Instance-level Dynamic Filter (IDF), and the process is described in Fig1. The PDF follows a spatial attention mechanism to capture pixel saliency and the IDF determines the direction of the dynamic weight vector adaptive to the input audio clip. The outputs of PDF and IDF are utilized to conduct the dynamic filter process.

\begin{figure}[t]
     \centering
     \includegraphics[scale=0.26]{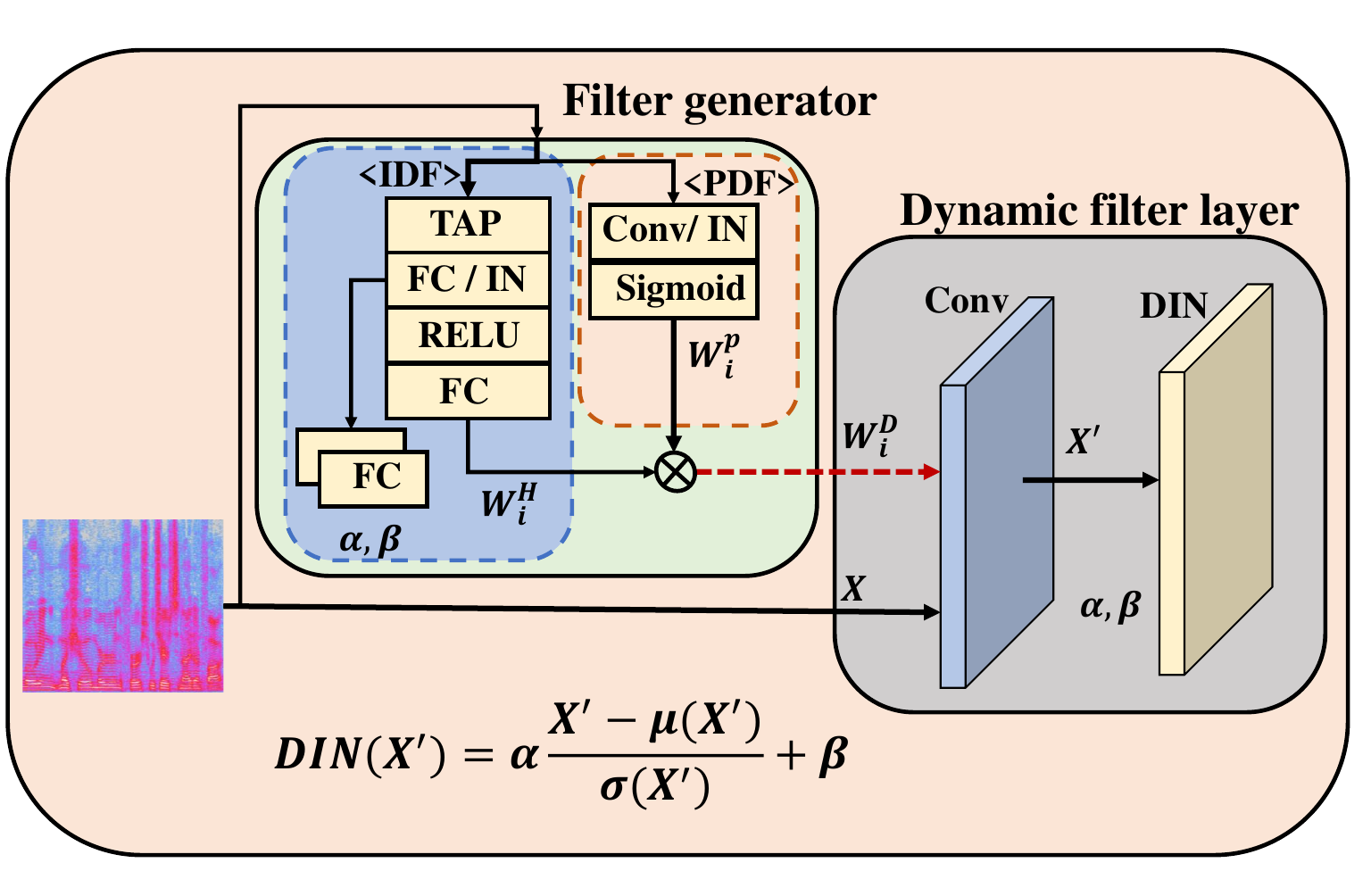}
     \caption{Pipeline of Lightweight Dynamic Convolution process: PDF denotes Pixel Dynamic Filter; IDF denotes Instance-level Dynamic Filter. N denotes total pixels in the T-F feature}
    \label{fig:three graphs}
\end{figure}
\section{Lightweight dynamic filter}
\subsection{Pixel Dynamic Filter}
The goal of PDF is to obtain an attention-based Time-Frequency (T-F) mask to measure a pixel level saliency. A single channel convolution is applied to T-F features ($x \in \mathcal{R}^{[T\times F]}$), and normalization with sigmoid activation is subsequently applied for mapping the feature scale from 0 to 1. The process of PDF is as follows: 
\begin{equation}
  w^P_i = \rho(IN(Conv(x_i, \bold{w}))),
  \label{eq2}
\end{equation}
where $Conv(x_i, \bold{w})$ denotes the convolution process with $x_i$ as the $i_{th}$ input data in mini-batch of the convolution and $\bold{w}$ represents the trainable weight vector ($\bold{w} \in \mathbb{R}^{[K\times K]}$, where $K$ denotes size of convolution kernel).  
Then Instance Normalization (IN) with sigmoid activation ($\rho(\cdot)$) are performed in series to obtain pixel level weights ( $w^P_i \in \mathbb{R}^{[T\times F,1]}$).
\subsection{Instance-level Dynamic Filter}
Conventional deep learning-based dynamic filters generate adaptive weights for the filter by learning them through the training process of a Convolutional Neural Network (CNN). While this approach has shown to be flexible and adaptive in various domains of diverse speaker characteristics or noise, it comes at the cost of high computational loads. 

As an efficient alternative to the convolution-based dynamic filter generation, we developed IDF which guides in capturing directional features of dynamic filter.
IDF is aimed to produce an audio representative weight vector, which is similar to speaker embedding in speaker verification. To this end, the temporally averaged feature ($H \in \mathbb{R}^{[F]}$) that preserves the frequency trait gets fed to an MLP-based model, and the output is used as vector direction of dynamic weight. The process of IDF is as follows:  
\begin{equation}
  \bold{w}^H = max(0, IN(H \cdot \bold{w_1} + \bold{b_1}))\cdot \bold{w_2} + \bold{b_2},
  \label{eq1}
\end{equation}
 where $\bold{w_i}$ and $\bold{b_i}$ are learning parameters. Two layers of MLP with IN and relu activation extract a weight vector ($\bold{w}^H \in \mathbb{R}^{[1,K \times K]}$). In this way, the weight vector is generated uniquely per input audio signal, and the dynamic filters ($\bold{w}^H$ and $w^P_i$) are multiplied to perform dynamic convolution. 

\subsection{Dynamic Convolution} 
The process of dynamic convolution is as follows:
\begin{equation}
  \bold{w}^D_i = w^P_i \odot \bold{w}^H, 
  \label{eq2}
\end{equation}
\begin{equation}
  x^{D}_{i} = IN(Conv(x_i ,\bold{w}^D_i)), 
  \label{eq2}
\end{equation}
\begin{equation}
  x^{'}_i = x^{D}_{i}+x_i, 
  \label{eq2}
\end{equation}
where $\bold{w}^D_i \in \mathbb{R}^{[T \times F, K \times K]}$ denotes dynamic weights for CNN computation. From equation 3, weight vector ($\bold{w}^H$) and scalar weight of pixel ($w^P_i$) perform element-wise product ($\odot$) to obtain the dynamic weights. Then, the convolution process using the dynamic weights is performed on the input and the result is normalized as in Equation 4. Through the skip connection, the normalized output ($x^{D}_i\in\mathbb{R}^{[T \times F]}$) is added to the input $x_i$ to be fed to the KWS model as shown in Equation 5. 
As such, the dynamic convolution-based filter generation adapts to each input effectively by enhancing features relevant to the KWS task. 
In conventional dynamic filters, a patch-level weight vector gets generated by a filter generator. Thus, the weight vector has different directions and magnitudes per patch. Our method, however, uniquely generates a unit directional vector for all the patches while the magnitude gets conditionally adjusted depending on each input patch. By decomposing the weight vector in terms of its direction and magnitude, the dimension of the weights becomes reduced significantly. This would significantly reduce the computational complexity of the dynamic filter. 
\subsection{Dynamic Instance Normalization}
Normalization technique is an important part to achieve promising KWS performance \cite{chang2021subspectral}. However, leveraging a conventional normalization process, where statistic weights (scale and bias) are used, might not efficiently handle blind environments. From the motivation of the previous study \cite{jing2020dynamic}, to address this, we leverage dynamic weights to implement feature normalization for the output of dynamic convolution. We firstly normalize the output of dynamic convolution by zero mean unit variance ($\mu_{x}$ and $\sigma_{x}$) and two different FC layers produce dynamic weights ($\alpha$ and $\beta\in \mathbb{R}^{[F]}$) respectively to adjust scale and bias. Here, an input of the linear layer is the output of the first linear layer in IDF. The process of Dynamic Instance Normalization (DIN) is as follows: 
 \begin{equation}
  DIN(x^{D}_{i}) =\alpha (x^{D}_{i} - \mu_{x^{D}_{i}})/\sigma_{x^{D}_{i}} +\beta,
  \label{eq1}
\end{equation}
 where $\mu_{x^{D}_{i}}$ and $\sigma_{x^{D}_{i}}$ denote mean and standard variation of the $x^{D}_{i}$. Instead of IN, DIN is applied to equation (4).
\subsection{Computational Complexity}

\begin{table}[]
  
  \label{tab:word_styles}
  \centering
    \caption{Comparison of the parameter numbers, computation time, and memory cost. Conv, DyConv, and LDyConv stand for static filter-based convolution, Dynamic filter convolution, and Lightweight Dynamic convolution respectively.}
  \begin{tabular}{|c|ccc|}
    \hline
    \textbf{Filter}& \textbf{Conv}& \textbf{DyConv}& \textbf{LDyConv}\\
    \hline
    Parameter& $K^2$&$K^4$ &$K^2(F+1)$\\
    Time& $O(K^2N)$&$O(K^4N)$ & $O(K^2N)$\\
    Space& $-$&$O(K^2N)$ &$O(N^2+K)$ \\
    \hline
  \end{tabular}
\end{table}
Our proposed dynamic convolution is applied to the front end of the KWS network for the dynamic feature extraction. A single-channel audio input gets fed to dynamic convolution and a single-channel output is employed as input for the KWS network. For this single-channel configuration, we compare computational costs (model parameters, time complexity, and space complexity) as summarized in Table 1. $N$ denotes the size of the pixel in the T-F feature which is equal to [$T \times F$]. For simplicity, we only compare the convolution process of the conventional Dynamic Convolution (DyConv) and our Lightweight Dynamic Convolution (LDyConv).\\
{\bf Model parameters.} In DyConv, the $K^2$ dimensional weight vector is driven by the patch size $K^2$. Thus, $K^4$ of parameters are used for DyConv implementation. In LDyConv, the PDF produces $K^2$ scalar weights from the patch dimension and the IDF generates a weight vector from $F$ by using two FC layers. Thus, LDyConv requires $K^2(F+1)$ parameters. Since $K^2$ and $F$ are of the same order, the number of parameters required for DyConv and LDyConv are similar.\\
{\bf Time complexity.} DyConv computes $K^4$ for every sample (pixel). With the total number of samples ($N$), DyConv has $O(K^4N)$ of time complexity. In LDyConv, the PDF computes $K^2$ for every sample while the IDF is only performed per audio clip. Thus, time complexity of LDyConv is $O(K^2(N+F))$ or it is approximately $O(K^2N)$ since $N>>F$. Our method would result in similar time complexity over the static filter.\\
{\bf Space complexity} As the DyConv generates patch-level dynamic weight, it has $O(K^2N)$ of space complexity for saving dynamic weight vector. On the other hand, our dynamic filter only requires $O(K^2+N)$ of space complexity.\\
In summary, our proposed lightweight dynamic convolution model has similar model complexity compared to a static convolution filter model. We confirm that our proposed model consumes 2K parameters and 220K Flops when $K=3$, $F=40$, and $T=98$.
\section{Experiment and discussion}
\subsection{Experimental Setup}
{\bf Dataset.} We used speech command datasets v1 and v2 \cite{warden2018speech} for evaluating the KWS performance. By following the DB guideline of the dataset, we utilized 10 keywords with two extra classes (unknown or silent) for model training, injected background noise, and added random time-shifting.\\
{\bf Noise setup.} For evaluating robustness against noise, we utilized DCASE \cite{mesaros2019acoustic}, Urbansound8K \cite{salamon2014dataset} and WHAM \cite{Wichern2019WHAM} datasets. These three datasets contain background noise of urban locations. For data augmentation, we randomly selected an audio sample from the noise data and mixed it with the test audio of speech command with 5 different Signal-to-Noise ratios (SNRs) [20dB, 15dB, 10dB, 5dB, and 0dB].\\
{\bf Acoustic feature extraction.} The acoustic feature we used is Mel Frequency Cepstral Coefficients (MFCC) constructed with 30ms of windows with 10ms overlap from an audio clip sampled at 16kHz. 64 Mel filters are employed to extract a Mel-spectrogram and 40 MFCC coefficients are extracted. This process gives a [40,98] size of audio features.\\
{\bf Computation setup.} All our experiments are done by using the Tensorflow deep learning package with RTX-2080 ti GPU. In the training process, we used a batch size of 100, 30K iterations, and an ADAM optimizer with a 0.001 initial learning rate. For every 10K iteration, the learning rate is decreased by 0.1.\\
{\bf Implementation detail.} In the PDF and the dynamic convolution process, we used $3\times3$ CNN kernel ($k=3$) dilated by (2,2) with a stride of 1. In the IDF, the first FC and second FC follow $40 \times 40$ and $40 \times k$ layer dimensions respectively. For DIN, two layers of FC which have $40 \times 40$ filter size respectively are utilized to produce $\alpha$ and $\beta$.
\subsection{Baselines}
Four different baseline architectures are used for comparisons. We implemented the proposed LDyConv on TENet architecture \cite{li2020small} and compared its performance with the following baseline models.\\
{\bf TCNet.} TCNet \cite{choi2019temporal} (or TC-Resnet) utilized temporal convolution and skip-connection for a fast and low computational cost model. TCNet8 contains 3 convolution blocks and 1 FC layer. Each convolution block has two layers of temporal convolution with a skip connection. Similarly, TCNet14 contains 6 convolution blocks and 1 FC layer. \\
{\bf TENet.} TENet \cite{li2020small} utilizes a depth-separable convolution framework. A convolution block contains three convolutions with batch normalization. TENet6 has 6 convolution blocks and 1 FC layer. Each convolution block has two 1D convolutions and 1 temporal convolution. TENet12 contains 12 convolution blocks and 1 FC layer. TENet has 32 output channels for each convolution block and TENet-n has 16 output channels for each convolution block.\\
{\bf MHA-RNN.} MHA-RNN\cite{rybakov2020streaming} utilizes CRNN and self-attention mechanism for the KWS model training. The output of CRNN feeds to a dot product-based Multi-Head Attention (MHA) model, and two layers of FC produce probability values for KWS.\\ 
{\bf Neural Architecture Search.} NAS is a network architecture designing method for deep learning applications and Differentiable Architecture Search (DARTS) is a variant of NAS that reduces search costs by weight sharing. We compared various state-of-art NAS methods in keyword spotting. Please see details of the model in \cite{mo2020neural,zhang2021autokws}.\\
{\bf BC-Resnet.} BroadCasted Resnet\cite{kim2021broadcasted} uses BC block which contains frequency and temporal depth-wise convolution with a SubSpectralNorm. In the BC block, the output of a 2D convolution is fed to the pooling layer and FC layer to represent audio features.\\
{\bf Lightweight convolution.} Lightweight convolution \cite{wu2019pay} (Lconv) is a separable convolution method by using weight sharing and weight normalization. A single block of Lconv contains two linear layers, Gated Linear Unit (GLU) activation, and lightweight convolution. we apply the single Lconv block at the front end of the TENet12 model. In the first linear layer, frequency is expended by 80, and GLU is carried out to the frequency dimension. Then, lightweight convolution ($H=10$) and the other linear layer are computed by preserving temporal and spectral features. Additionally, by following \cite{wu2019pay}, we perform Dynamic convolution (Dyconv). Instead of employing the static weight in the lightweight convolution, a single linear layer is used to produce weights for Lconv. In our implementation, the Lconv block requires 982K of Flops. and 4.9K of parameters. In the Dconv, 1373K of Flops. and 6.6K of parameters are used. 
 \subsection{Result discussion}

\begin{table}[t]
  \centering
\scriptsize
    \caption{Comparison with lightweight models on Speech Command v1 and v2: Par. and Flops. denote Model parameters and computational cost respectively. Notation of $\dagger$ denotes the application of Spec-Augmentation \cite{park2019specaugment}. For an accurate experiment, 8 times averaging accuracy and best performance are presented.}
  \label{tab:word_styles}
  \begin{tabular}{|c|c|cc|cc|}
    \hline
    \multicolumn{1}{|c|}{\multirow{2}{*}{\textbf{Model}}}&\multicolumn{1}{c|}{\multirow{2}{*}{\textbf{(Par.,Flops.)}}}&\multicolumn{2}{c|}{V1}&\multicolumn{2}{c|}{V2}\\\cline{3-6}
    \multicolumn{1}{|c|}{}&\multicolumn{1}{c|}{}&\textbf{Acc}& \textbf{Best}& \textbf{Acc}& \textbf{Best}\\
    \hline
    TCNet8\cite{choi2019temporal}& (145K,4.40M)& - &96.2&- &- \\
    TCNet14\cite{choi2019temporal}& (305K,8.26M)& - &96.6&96.53&96.8 \\
    \hline
    TENet12\cite{li2020small}& (100K,6.42M)& - &96.6 &97.10&97.3\\
    TENet12$^{\dagger}$\cite{li2020small}& (100K,6.42M)& 97.19 &97.3 &97.43&97.6\\
    \hline
    MHA-RNN$^{\dagger}$\cite{rybakov2020streaming}& (743K,87.2M)& - &97.2& -&98.0 \\
    \hline
    BC-ResNet3$^{\dagger}$\cite{kim2021broadcasted}& (54.2K,32.4M)& 97.6 & - &98.2&-\\
    BC-ResNet6$^{\dagger}$\cite{kim2021broadcasted}& (188K,106.2M)& 97.9 & - &98.6&-\\
    BC-ResNet8$^{\dagger}$\cite{kim2021broadcasted}& (321K,178.2M)& 98.0 & - &98.7&-\\
    \hline
    NAS2\cite{mo2020neural}& (886K,-)& - &97.2& -& -\\
    Random\cite{zhang2021autokws}& (196K,8.8M)& 96.58 &96.8& -&- \\
    DARTS\cite{zhang2021autokws}& (93K,4.9M)& 96.63 &96.9& 96.92&97.1 \\
    F-DARTS\cite{zhang2021autokws}& (188K,10.6M)& 96.70&96.9 &97.1& 97.4 \\
    N-DARTS\cite{zhang2021autokws}& (109K,6.3M)& 96.79 &97.2& 97.18&97.4 \\
    \hline
    Lconv\cite{wu2019pay}&(105K,7.40M)&96.88&97.0& 97.24&97.3\\
    Dconv\cite{wu2019pay}&(107K,7.69M)&96.89&97.1& 96.26&97.4\\
    \hline
    LDy-TENet12 \tiny{(w/o DIN)}& (102K,6.64M)& 96.95 &97.1 & 97.35&97.6\\
    LDy-TENet12 \tiny{(w/o DIN)}$^{\dagger}$& - & 97.42 &97.6 & 97.66&97.7\\
    LDy-TENet12& (105K,6.97M)& 96.94 &97.1 & 97.40&97.6\\
    LDy-TENet12$^{\dagger}$& - & 97.43 &97.6 & 97.67&97.7\\
    
    \hline
  \end{tabular}
\end{table}

\begin{table}[h]
  \label{tab:word_styles}
  \centering
    \caption{Comparison with Unseen noise environment on Speech Command v1: experiment is performed on LDy-TENet12, TENet12, Lconv and Dconv models.}
  \scriptsize
\begin{tabular}{|c|c|ccccc|}
\hline
\multicolumn{1}{|c|}{\multirow{3}{*}{\textbf{Noise}}} & \multicolumn{1}{c|}{\multirow{3}{*}{\begin{tabular}[c]{@{}c@{}}SNR\\ (dB)\end{tabular}}} & \multicolumn{5}{c|}{\textbf{Model}}\\ \cline{3-7}
\multicolumn{1}{|c|}{}&\multicolumn{1}{c|}{}&\multicolumn{2}{|c|}{\textbf{LDy.}}&\multicolumn{1}{|c|}{\multirow{2}{*}{\textbf{Dconv}}}&\multicolumn{1}{|c|}
{\multirow{2}{*}{\textbf{Lconv}}}& \multicolumn{1}{|c|}{\multirow{2}{*}{\textbf{TENet}}}\\\cline{3-4}
\multicolumn{1}{|c|}{}&\multicolumn{1}{c|}{}&\textbf{\tiny{w DIN}}&\textbf{\tiny{w/o DIN}}&\multicolumn{1}{|c|}{}&\multicolumn{1}{c|}{}&\multicolumn{1}{|c|}{}\\\hline

\multirow{5}{*}{{\begin{tabular}[c]{@{}c@{}}\textbf{DCASE}\end{tabular}}}  
                                &\textbf{20}&97.08&\textbf{97.17} & 96.90 &97.07&97.10\\
						         &\textbf{15}&96.87&\textbf{96.91}& 96.76&96.80&96.84\\
					            &\textbf{10}&96.02&\textbf{95.98}& 95.73&95.92&95.77\\
                                 &\textbf{5}&\textbf{94.38}&94.15& 93.79&94.11&93.85\\
					            &\textbf{0}&\textbf{90.64}&90.42& 89.18&89.53&89.20\\\hline
\multirow{5}{*}{\begin{tabular}[c]{@{}c@{}}\textbf{Urban}\end{tabular}}   
                                &\textbf{20}&\textbf{96.36}&96.34& 96.27&96.25&96.34\\
						         &\textbf{15}&\textbf{95.50}&\textbf{95.50}& 95.42&95.42&95.49\\
					            &\textbf{10}&93.99&94.15&93.74&93.76&93.49\\
                            &\textbf{5}&\textbf{91.11}&90.80& 90.02&90.03&90.17\\
					            &\textbf{0}&\textbf{82.32}&81.39& 79.73&80.00&80.26\\\hline
\multirow{5}{*}{\begin{tabular}[c]{@{}c@{}}\textbf{WHAM}\end{tabular}} 
                                &\textbf{20}&\textbf{96.67}&96.66& 96.51& 96.61&96.60\\
						         &\textbf{15}&\textbf{95.97}&95.93& 95.86& 95.94&\textbf{95.98}\\
					            &\textbf{10}&93.64&\textbf{93.75}& 93.33& 93.73&93.51\\
                                &\textbf{5}&89.74&\textbf{89.93}& 89.49& 90.11&89.32\\
					            &\textbf{0}&\textbf{79.73}&78.86& 78.32& 79.13&78.39\\
\hline
\end{tabular}
\end{table}
\begin{table}[h]
  \label{tab:word_styles}
  \centering
    \caption{Comparison with Unseen noise environment on Speech Command v2.}
  \scriptsize
\begin{tabular}{|c|c|ccccc|}
\hline
\multicolumn{1}{|c|}{\multirow{3}{*}{\textbf{Noise}}} & \multicolumn{1}{c|}{\multirow{3}{*}{\begin{tabular}[c]{@{}c@{}}SNR\\ (dB)\end{tabular}}} & \multicolumn{5}{c|}{\textbf{Model}}\\ \cline{3-7}
\multicolumn{1}{|c|}{}&\multicolumn{1}{c|}{}&\multicolumn{2}{|c|}{\textbf{LDy.}}&\multicolumn{1}{|c|}{\multirow{2}{*}{\textbf{Dconv}}}&\multicolumn{1}{|c|}
{\multirow{2}{*}{\textbf{Lconv}}}& \multicolumn{1}{|c|}{\multirow{2}{*}{\textbf{TENet}}}\\\cline{3-4}
\multicolumn{1}{|c|}{}&\multicolumn{1}{c|}{}&\textbf{\tiny{w DIN}}&\textbf{\tiny{w/o DIN}}&\multicolumn{1}{|c|}{}&\multicolumn{1}{c|}{}&\multicolumn{1}{|c|}{}\\\hline
\multirow{5}{*}{{\begin{tabular}[c]{@{}c@{}}\textbf{DCASE}\end{tabular}}}  
                                &\textbf{20}&\textbf{97.24}&97.01& 96.80&96.94&96.79\\
						         &\textbf{15}&\textbf{96.74}&96.35 & 96.24 &96.29&96.05\\
					            &\textbf{10}&\textbf{96.06}&95.49& 95.28&95.23&95.18\\
                                 &\textbf{5}&\textbf{94.22}&93.74& 92.95 &92.85&92.82\\
					            &\textbf{0}&\textbf{90.08}&89.05& 87.71 &87.66&87.36\\\hline
\multirow{5}{*}{\begin{tabular}[c]{@{}c@{}}\textbf{Urban}\end{tabular}}   
                                &\textbf{20}&\textbf{96.55}&96.22& 96.04 &96.16&96.06\\
						         &\textbf{15}&\textbf{95.20}&94.99& 94.44&94.55&94.65\\
					            &\textbf{10}&\textbf{93.80}&93.45&92.61&92.41&92.67\\
                            &\textbf{5}&\textbf{89.28}&88.45& 87.60 &87.18&87.51\\
					            &\textbf{0}&\textbf{81.77}&80.25 & 78.39 &78.16&78.24\\\hline
\multirow{5}{*}{\begin{tabular}[c]{@{}c@{}}\textbf{WHAM}\end{tabular}} 
                                &\textbf{20}&\textbf{96.37}&96.11& 96.00& 96.19&96.06\\
						         &\textbf{15}&\textbf{95.72}&95.17& 94.87& 94.88 &94.99\\
					            &\textbf{10}&\textbf{93.57}&93.38 & 93.00& 93.15&92.89\\
                                &\textbf{5}&\textbf{88.91}&88.46& 87.99& 87.83&87.41\\
					            &\textbf{0}&\textbf{78.66}&77.43& 76.31& 76.65&75.93\\
\hline
\end{tabular}
\end{table}
Tables 2, 3, and 4 summarize the KWS results on the Speech Command dataset v1 and v2. Our proposed method, Lconv block, and Dconv block are applied to the front end of the TENet12 model. For a more thorough model evaluation, we repeated the experiment 8 times.\\
{\bf Small footprint KWS.} Table 2 shows small-footprint KWS performances over state-of-the-art methods. We compare the learning parameters, Flops., averaging accuracy, and the highest accuracy over the 8 repeated experiments. For the fair comparison, we perform Spec-Augmentation \cite{park2019specaugment} during the training, and the results are indicated by $\dagger$. From the results, we confirm that our methods (LDy-TENet) show improved performance over TENet based model. Particularly, the LDy-TENet6-n shows similar performance over the TENet12 which is a 3 times larger model. Compared with the NAS-based models, our LDy-TENet12 achieves the best-averaging accuracy with low computational costs. The performance improvement is not significant between the TENet12 and the lightweight convolution models (Lcon and Donv). Additionally, the parameters and Flops. of Lconv and Dconv are higher than our method, while they take degraded performance. Although BC-Resnet-based models show improved performance over our method, these models require high computational resources (FLOPS.) for their implementation. On the other hand, our method can be worked by using small FLOPS.\\
{\bf Unseen noise environment.} Tables 3 and 4 summarize KWS results on the unseen noisy environment with 3 different noise datasets and 5 different SNRs.
The results of the Lconv and Dconv are less significant since they show similar performance with the TENet. On the other hand, our proposed model shows more robust performance over the baselines, and particularly the performance is improved when the SNR is low. Compared with the TENet, 2.1\% (v1) and 3.5\% (v2) performance improvements are shown in the Urban 0dB condition. Especially, leveraging DIN enhances the robustness of the model in the blind environment. DIN delivers performance improvements when 5 and 0 dB SNR environments\\
As a result, our proposed dynamic filter in the front of the network would enhance the performance of KWS in unseen noisy environments. Especially, as the model takes two main parts (PDF and IDF), it can be implemented with a small computational cost.

\section{Conclusion}
The main focus of this study was to develop a lightweight dynamic filter model for an acoustic feature extractor in Keyword spotting. We proposed the Lightweight Dynamic Convolution model which decomposes a dynamic filter into two parts (pixel and kernel) for alleviating the issues of computational cost and noise robustness. In addition, Dynamic Instance Normalization delivers performance improvement over noisy environments. The process has a small footprint and through the relevant experiments, it is shown fast compared to the conventional dynamic convolution method while retaining the adaptability of a dynamic filter. The experiments confirmed that our proposed lightweight model is robust on unseen noise over lightweight models.
\bibliographystyle{IEEEtran}
\bibliography{strings}
\end{document}